\documentclass[preprint2]{aastex}
\def\stacksymbols #1#2#3#4{\def\theguybelow{#2}
        \def\verticalposition{\lower#3pt}
        \def\spacingwithinsymbol{\baselineskip0pt\lineskip#4pt}
        \mathrel{\mathpalette\intermediary#1}}
\def\intermediary #1#2{\verticalposition\vbox{\spacingwithinsymbol
        \everycr={}\tabskip0pt
        \halign{$\mathsurround0pt#1\hfil##\hfil$\crcr#2\crcr
                \theguybelow\crcr}}}
\def\lta{\stacksymbols{<}{\sim}{2.5}{.2}}
\def\gta{\stacksymbols{>}{\sim}{3}{.5}}

\begin{document}

\title{EVOLUTION OF GAS IN ELLIPTICAL GALAXIES}

\author{William G. Mathews$^1$ and Fabrizio Brighenti$^{1,2}$}

\affil{$^1$University of California Observatories/Lick Observatory,
Board of Studies in Astronomy and Astrophysics,
University of California, Santa Cruz, CA 95064\\
mathews@lick.ucsc.edu}

\affil{$^2$Dipartimento di Astronomia,
Universit\`a di Bologna,
via Ranzani 1,
Bologna 40127, Italy\\
brighenti@bo.astro.it}

\begin{abstract}
We review the origin and structure of hot (cooling flow) gas in
elliptical galaxies.  X-ray observations can be used to determine the
stellar mass to light ratio, the mass profiles of dark matter halos,
and the interstellar magnetic field.  Interstellar gas cools over a
large volume, forming stars with a bottom-heavy IMF.  For consistency
with the thin fundamental plane, young stars must be optically
luminous.  Circular X-ray isophotes in rotating elliptical galaxies
indicate distributed radiative cooling or strong interstellar
turbulence.
\end{abstract}


\vskip.2in

\section{Origin of the Hot Gas}
The ratio of gas to stellar mass within giant elliptical galaxies 
is comparable to that in spiral galaxies, but the gas is much hotter,
$T \sim 10^7$ K.
Inside the half light radius ($r_e \sim 10$ kpc) most of the 
hot interstellar gas 
is produced by mass loss from the evolving old stellar population. 
Gas beyond $r_e$ is also supplied by an inflow of local 
intergalactic gas which shocks to the virial temperature 
$T_{vir} \sim 10^7$ K. 
This more weakly bound gas can be stripped by ram pressure 
in rich clusters or either increased or depleted 
by tidal exchanges among galaxies in groups. 
The dominant stellar population in elliptical galaxies is 
typically very old, and the dissipationless 
assembly of the stars into the 
$r^{1/4}$ configurations observed today probably also occurred 
in the distant past, $z \gta 1$.
X-ray emission from the interstellar gas,
often referred to as a ``cooling flow'', indicates that the gas is 
losing energy.
Paradoxically, however, the gas does not cool as it loses 
energy since 
it is immediately heated by 
$Pdv$ compression in the galactic gravitational potential 
as the gas flows subsonically inward. 
The modest interstellar iron abundance ($z_{Fe} \sim 0.5 - 1$ solar) 
in $r \lta r_e$, mostly due to Type Ia supernovae, 
indicates that supernova heating is not very important.
The radial temperature gradient in cooling flows is 
small, $T \sim T_{vir}$ throughout.

It is easy to estimate 
the total interior mass $M(r)$ in elliptical galaxies 
by entering the observed hot gas density and temperature 
profiles into the equation of hydrostatic equilibrium.
For several well-observed elliptical galaxies in Virgo 
the indicated mass in the range $0.1 r_e \lta r \lta r_e$ 
is equal to the stellar mass, i.e., the 
(dynamically determined) central stellar mass to 
light ratio is accurate and 
fairly constant with galactic radius until 
the dark halo dominates beyond $r_e$
(Brighenti \& Mathews 1997).
In the inner regions, $r \lta 0.1r_e$, however, the total mass 
indicated from the X-rays is less than the known stellar 
mass.
This peculiar result can be understood if the hot gas there is 
being supported by some additional pressure, not included in the 
hydrostatic equation.
The most likely source of additional 
pressure support is magnetic 
with strength $B \sim 100\mu$G. 
Such central fields can arise quite 
naturally from turbulent amplification of stellar seed fields
(Mathews \& Brighenti 1997). 
The presence of these large fields may be completely 
independent of past or present AGN nuclear activity.

Simple gas dynamical models for the origin and evolution of 
hot gas in ellipticals have been quite successful 
(Brighenti \& Mathews 1999a,b).
The models begin with a tophat cosmological perturbation 
that develops into a Navarro-Frenk-White (NFW)
dark halo growing in mass from the inside out. 
The baryons flow toward this perturbation, shock 
to $\sim T_{vir}$
and begin to lose energy by X-ray emission.
At some early time, around age  $\sim 1$ Gyr, 
stars are assumed to form with a Salpeter IMF, immediately 
releasing Type II supernova energy and metal enrichment.
This starburst drives a strong shock wave upstream into 
the converging cosmic gas.
At a slightly later time, $\sim 2$ Gyrs, 
after enough baryons have entered the halo, the 
de Vaucouleurs profile for the old stars is formed, 
simulating the merger process. 
From that time 
forward we solve the detailed gas dynamical equations, 
including stellar mass loss and Type Ia supernovae 
consistent with the observed rate.
When the cooling flow gas is evolved to the present time, 
with continued inflow of gas from the local 
intergalactic environment,
we find very good agreement with the observed hot gas 
density, temperature and iron abundance profiles 
in $r \gta 0.1r_e$. 
The entropy and gas mass fraction within the hot 
gas are accurately reproduced in this calculation. 

\section{Where Is the Cooled Gas and How Did it Cool?}

One of the great cooling flow 
mysteries, both on galactic and galaxy
cluster scales, is the uncertain fate of cooled gas.
For massive elliptical galaxies we know that the cooling flow 
cannot proceed all the way to the very center before 
cooling, even if angular momentum is ignored, because 
the masses of gas that cool ($\sim 4 \times 10^{10}$ 
$M_{\odot}$) far exceed the masses of 
the central black holes observed in elliptical galaxies.
Furthermore, VLA and other radio observations have 
set upper limits on the mass of HI and H$_2$ gas 
that are ludicrously small in many massive elliptical 
galaxies, $M(HI) + M(H_2) < 10^7$
$M_{\odot}$, so this endstate is not an option. 
Optical studies show that the cooled gas 
cannot be photoionized, 
typically $M(HII) \sim 10^5$ $M_{\odot}$ 
which is much too small.
However, the universal evidence for HII gas 
in cooling flows with intermediate 
temperatures, $T \sim 10^4$K, 
is clear evidence that the hot gas is indeed cooling 
over an extended region.

Even if the cooled gas is difficult to observe directly, 
its mass must contribute in a measurable way. 
X-ray data indicates that mass of old stars 
is sufficient to account for all mass 
in $r \lta r_e$.
The best solution of this problem is to assume that 
the hot interstellar gas cools and deposits its 
mass over a large galactic volume 
within $r_e$, converting its mass rapidly into 
young stars
(Brighenti \& Mathews 2000a).
It seems plausible that this cooling mass ``dropout'' 
is concentrated toward the center of the flow 
where the interstellar gas density and the radiation 
emissivity are greatest.
Since there is no optical evidence for 
massive young stars in most elliptical galaxies, 
many authors have supposed that only 
very low mass stars formed from the cooled gas. 
However, unless the dropout profile 
is carefully orchestrated,
dark baryonic stellar matter 
would interfere with the constancy of the dynamic 
mass to light ratio in $0.1r_e \lta r \lta r_e$ 
discussed above.
An additional young population of 
low mass, optically dark stars 
would also cause a large, measurable scatter perpendicular 
to the 
fundamental plane even if elliptical galaxies 
were otherwise structurally and dynamically homologous
(Mathews \& Brighenti 2000).
The best means of avoiding these difficulties is 
to assume that the continuously forming dropout 
stellar population is optically luminous  
with a mass to light ratio 
that does not differ greatly from that of the old
dominant stellar population. 
It is possible to imagine how this could happen. 
The interstellar pressure in bright elliptical 
galaxies is about $10^4$ times larger than in the 
Milky Way disk. 
As a consequence, very small masses of cooled 
HI or H$_2$ gas,  
$M \sim 2$ $M_{\odot}$ 
(weakly heated and ionized by X-rays;
Mathews \& Brighenti 1999),   
become Jeans unstable and collapse.
If $\sim 7$ percent of the stars in large ellipticals
have continuously formed with a Salpeter slope extending
only to $\sim 2$ $M_{\odot}$, their contribution to 
the stellar H$\beta$ index could explain the 
apparent youthful age of the stellar spectrum 
observed in many massive elliptical galaxies.

Spatially extended 
star formation in elliptical galaxies is obviously 
a difficult problem since 
the gas is about $10^6$ times 
hotter than temperatures 
in typical star forming regions in the Milky Way.
It has generally been assumed that localized 
thermal instabilities 
play an important role in concentrating 
the gas until gravity dominates.
Thermal instabilities traditionally develop 
from small low entropy regions of enhanced 
radiative emission where the gas density 
is slightly larger and the temperature is less 
than surrounding hot gas at the same pressure.
Remarkably, Loewenstein (1989) showed that 
low entropy inhomogeneities oscillate radially 
in a cooling flow atmosphere 
and on average cool no faster than the surrounding 
flow, provided the perturbed regions move 
without drag. 
However, in reality 
the strong drag interaction with the ambient gas completely 
damps the oscillations.
It now seems likely that small enhancements in the 
local interstellar magnetic field, not the entropy, may 
initiate cooling toward star formation. 
Magnetically buoyant regions can float at 
nearly a fixed radius in the cooling flow until 
the internal gas cools by radiative losses. 
In these floating regions, where most of the pressure 
is magnetic, 
radiative cooling proceeds at nearly constant gas 
density until the gas recombines and cools 
to star-forming temperatures 
(Mathews \& Brighenti 1999; 2000).
This model for localized cooling far from the 
centers of cooling flows is consistent 
with the recent discovery of 
spatially extended, cooler X-ray absorbing gas 
(Sanders, Fabian \& Allen 2000; Buote 2000a,b). 

\section{Are Cooling Flows Really Heating Flows?}

The uncertain physics of inhomogeneous cooling regions 
and low mass star formation 
in elliptical galaxies has led some authors to 
suggest that the cooling flows are not cooling at all, 
but are being heated by the release of energy from 
an active nucleus.
Ciotti \& Ostriker (1997; 2000) have proposed a model 
in which interstellar gas in all elliptical galaxies 
is intermittently and explosively 
Compton heated to $T_c \sim 10^9$ K by intense
bursts of radiation from a central active nucleus,
powered by a small amount of interstellar
gas that cools onto the central black hole.
Although the Compton temperature of observed AGN 
and quasar continua 
is very much less, $T_c \lta 10^7$K, 
it would be difficult to observe the much harder 
Ciotti-Ostriker radiation 
continuum from elliptical galaxies if the time scale 
for the radiation bursts is sufficiently short. 
Binney \& Tabor (1995) propose a
similar model in which
the accretion of a small amount of centrally
cooled gas by
the central hole leads to the production of 
violent radio
jets and an expanding relativistic plasma that
shocks and heats the inner cooling flow gas sufficiently to
offset its energy losses by thermal radiation.
Later, as the energy of the relativistic plasma decays,
the cooling flow is briefly re-established,
more gas cools onto the central hole
and the AGN radio lobe heating process repeats again.
By this means it is argued that
radiative cooling can be balanced by
central heating so no distributed mass dropout
is required.
However, recent {\it Chandra} observations of the hot
(cooling flow) gas
surrounding the radio lobes of Hydra A
(McNamara et al. 2000) 
and Perseus A (Fabian et al. 2000) show no evidence
that the radio sources are heating the gas. 
Instead, the radio lobes appear to be pushing the hot gas 
aside, gently and adiabatically.

\section{Why are Cooling Flow X-ray Contours so Round?}

While the stellar systems in 
most X-ray luminous elliptical galaxies 
are not rotatationally flattened, they 
do in fact rotate at $v \sim 50 - 100$ km s$^{-1}$. 
Gas lost from the rotating stellar system  
is rapidly heated to the virial temperature and   
must initially share 
the mean local stellar angular momentum.
If this gas conserves angular momentum 
as it moves slowly inward with the cooling flow, 
it would spin up to the circular velocity
$v_{circ} \sim 400$ km s$^{-1}$, and settle onto a 
thin cold disk of size $r_d \gta r_e$.
Remarkably, 
the flattening of X-ray isophotes that would accompany 
such a flow has never been observed, instead 
the inner X-ray contours are almost perfectly round
(Hanlan \& Bregman 2000).
There are at least three explanations for this 
peculiar result (Brighenti \& Mathews 2000b). 
Perhaps the most natural explanation is 
that angular momentum is lost by mass dropout 
in the cooling process. 
Alternatively, strong interstellar turbulence could 
transfer angular momentum to the outer regions of the 
cooling flow, but the turbulent energy density required 
exceeds that of known sources.
Finally, torques from strong central magnetic fields, 
discussed above, may be sufficient to transport 
angular momentum away from the inner cooling flows.

\end{document}